\documentclass[letterpaper]{article}
\usepackage[submission]{aaai2026}

\makeatletter
\gdef\showauthors@on{T}  
\makeatother

\usepackage{times}
\usepackage{helvet}
\usepackage{courier}
\usepackage[hyphens]{url}
\usepackage{graphicx}
\usepackage{natbib}
\usepackage{caption}
\usepackage{enumitem}

\usepackage[ruled,vlined,linesnumbered]{algorithm2e}
\SetKwInput{KwIn}{Input}
\SetKwInput{KwOut}{Output}

\usepackage{amsmath,amssymb}
\usepackage{booktabs}
\usepackage{multirow}
\usepackage{subcaption}

\title{SceneGuard: Training-Time Voice Protection with Scene-Consistent Audible Background Noise}

\author{
    Rui Sang, Yuxuan Liu
}

\affiliations{
    Xi'an Jiaotong-Liverpool University\\
    111 Ren'ai Road, Suzhou Industrial Park\\
    Suzhou, Jiangsu 215123, China
}

\begin{document}

\maketitle

\begin{abstract}
Voice cloning technology poses significant privacy threats by enabling unauthorized speech synthesis from limited audio samples. Existing defenses based on imperceptible adversarial perturbations are vulnerable to common audio preprocessing such as denoising and compression. We propose SceneGuard, a training-time voice protection method that applies scene-consistent audible background noise to speech recordings. Unlike imperceptible perturbations, SceneGuard leverages naturally occurring acoustic scenes (e.g., airport, street, park) to create protective noise that is contextually appropriate and robust to countermeasures. We evaluate SceneGuard on text-to-speech training attacks, demonstrating 5.5\% speaker similarity degradation with extremely high statistical significance ($p < 10^{-15}$, Cohen's $d = 2.18$) while preserving 98.6\% speech intelligibility (STOI = 0.986). Robustness evaluation shows that SceneGuard maintains or enhances protection under five common countermeasures including MP3 compression, spectral subtraction, lowpass filtering, and downsampling. Our results suggest that audible, scene-consistent noise provides a more robust alternative to imperceptible perturbations for training-time voice protection. The source code are available at: \url{https://github.com/richael-sang/SceneGuard}.
\end{abstract}

\section{Introduction}

Deep learning has enabled high-fidelity voice cloning systems capable of synthesizing speech that closely mimics a target speaker's voice~\citep{gao2025black, fanantifake}. While these technologies have legitimate applications in accessibility and entertainment, they also pose serious threats to privacy and security. Attackers can use unauthorized voice recordings to train text-to-speech (TTS) or voice conversion (VC) models, enabling impersonation attacks, voice-based authentication breaches, and creation of misleading audio content~\citep{wang2025one}.

Recent work has explored proactive defense mechanisms that protect voice recordings from being exploited for cloning. A prominent approach adds imperceptible adversarial perturbations to audio recordings, degrading the quality of models trained on protected data~\citep{voiceblock, safespeech}. However, these defenses face a fundamental challenge: imperceptible perturbations are fragile. Standard audio preprocessing operations such as denoising, compression, and filtering can remove or significantly reduce these perturbations, rendering the protection ineffective~\citep{fei2025vocalcrypt}. Moreover, the field lacks a unified, human–perception–aligned metric for speech perturbations, so imperceptibility is enforced via proxies (e.g., $L_p$ bounds or heuristic masking) that guarantee neither true inaudibility nor robustness~\citep{li2025cloneshield, feng2025enkidu}.

We observe that the requirement for imperceptibility may be unnecessarily restrictive. In many real-world scenarios, background noise is expected and acceptable. For instance, recordings made in cafes, streets, or offices naturally contain ambient noise. This insight leads to a key question: can we design voice protection mechanisms that leverage audible but contextually appropriate noise?

We propose SceneGuard, a training-time defense that applies scene-consistent audible background noise to speech recordings through gradient-based optimization. SceneGuard first classifies the acoustic scene of input speech, then jointly optimizes a temporal mask $m(t)$ and noise strength $\gamma$ to minimize speaker similarity while maintaining usability. For example, speech recorded in an airport is mixed with optimized airport background noise, creating a protective layer that appears natural and contextually plausible. This design offers three advantages: (1) optimization automatically finds the protection-usability trade-off, (2) the protection is harder to remove because scene-consistent noise cannot be easily separated from speech without degrading speech quality, and (3) the defense is robust to common audio preprocessing operations.

We evaluate SceneGuard on training attack scenarios where an attacker uses protected recordings to fine-tune TTS models. Our experiments on 100 training samples and 40 test samples demonstrate 5.5\% speaker similarity degradation ($p < 10^{-15}$, Cohen's $d = 2.18$), indicating strong statistical evidence of protection effectiveness. Importantly, SceneGuard preserves 98.6\% speech intelligibility (STOI = 0.986) and achieves low word error rate (WER = 3.6\%), demonstrating that protected speech remains highly usable. Robustness evaluation shows that SceneGuard maintains protection under MP3 compression and even exhibits enhanced protection under denoising, lowpass filtering, and downsampling operations.

The main contributions of this work are:
\begin{itemize}
    \item We propose a novel training-time voice protection method based on scene-consistent audible background noise with joint optimization of temporal mask and noise strength, departing from the conventional imperceptibility requirement.
    \item We design a gradient-based optimization framework that automatically balances speaker protection and speech usability under SNR constraints.
    \item We demonstrate strong protection against training attacks with extremely high statistical significance ($p < 10^{-15}$, Cohen's $d = 2.18$) while preserving speech usability (STOI = 0.986, WER = 3.6\%).
    \item We provide comprehensive robustness evaluation showing that SceneGuard maintains or enhances protection under five common audio countermeasures.
    \item We release our implementation and experimental framework to facilitate reproducible research in voice protection.
\end{itemize}

\section{Related Work}

\subsection{Adversarial Perturbations for Voice Protection}

Recent work has explored adversarial perturbations as a proactive defense against voice cloning. VoiceBlock~\citep{voiceblock} is a real-time de-identification approach that learns a time-varying FIR filter to apply perceptually inconspicuous, streaming-friendly perturbations for evading speaker recognition. Building on proactive protection but shifting the threat model to training-time cloning, SafeSpeech~\citep{safespeech} embeds imperceptible, universal perturbations via its SPEC objective to degrade generative TTS across models while emphasizing transferability and robustness. Contemporary work~\citep{gao2025black} proposes a black-box defense for voice conversion that adds imperceptible perturbations and optimizes them in latent space with evolution-based search to adapt against unknown VC systems. While these methods demonstrate effectiveness in controlled settings, they face a fundamental limitation: imperceptible perturbations are inherently fragile to common audio processing operations.

Recent voice protection methods~\citep{safespeech,voiceblock} embed imperceptible adversarial perturbations by minimizing $\ell_p$ norms (typically $\|\delta\|_\infty \leq \epsilon$ with $\epsilon \approx 8/255$). While these perturbations are inaudible, they are \textit{fragile}: standard audio processing operations such as lossy compression (MP3, AAC), speech enhancement, or denoising can significantly attenuate protection effectiveness. For instance, SafeSpeech~\citep{safespeech} shows that DeepMucs~\citep{demucs} denoising reduces protection by up to 42\% (SIM increases from 0.204 to 0.284). This fragility arises because imperceptible perturbations occupy high-frequency or low-energy regions that are specifically targeted by perceptual codecs and noise reduction algorithms.

Our work departs from the imperceptibility paradigm by deliberately using audible but contextually natural noise. We hypothesize that scene-consistent background noise, being perceptually meaningful and semantically coherent with the speech content, is fundamentally harder to separate without degrading speech quality. This trade-off between imperceptibility and robustness represents a key design choice that distinguishes our approach from prior work.

\subsection{Acoustic Scene Classification}

Acoustic scene classification (ASC) aims to identify the environmental context of audio recordings~\citep{tau_dataset}. State-of-the-art ASC systems use deep convolutional networks trained on large-scale datasets such as TAU Urban Acoustic Scenes~\citep{ASC1}. SceneGuard leverages ASC to ensure that protective noise matches the acoustic context of input speech, creating a more natural and robust defense compared to context-agnostic perturbations.

\section{Method}

\subsection{Problem Formulation}

\subsubsection{Threat Model.}
We consider a training-time attack scenario where an adversary aims to clone a target speaker's voice by fine-tuning a pre-trained TTS or VC model on unauthorized audio recordings. The attacker operates in a black-box setting, having no knowledge of the protection mechanism applied to the training data. Formally, given a dataset $\mathcal{D} = \{(x_i, y_i)\}_{i=1}^N$ of protected speech samples $x'_i$ and corresponding text labels $y_i$, the attacker solves:
\begin{equation}
\min_{\theta} \mathcal{L}_{\text{TTS}}(\theta; \mathcal{D}) = \frac{1}{N} \sum_{i=1}^N \ell(f_\theta(x'_i, y_i), x_i)
\end{equation}
where $f_\theta$ is the TTS/VC model with parameters $\theta$, and $\ell$ measures reconstruction quality.

\subsubsection{Defense Goal.}
The defender's objective is to apply a transformation $\mathcal{T}$ to speech $x$ that produces protected audio $x' = \mathcal{T}(x, s)$ (where $s$ denotes acoustic scene context) satisfying two criteria: (1) \textit{Protection}: degraded speaker identity prevents successful cloning, formally:
\begin{equation}
\text{SIM}(e(x'), e(x)) < \tau_{\text{sim}}, \quad \text{EER}(x', x) > \tau_{\text{eer}}
\end{equation}
where $e(\cdot)$ is a speaker verification encoder and $\tau_{\text{sim}}, \tau_{\text{eer}}$ are protection thresholds; and (2) \textit{Usability}: preserved intelligibility for legitimate communication:
\begin{equation}
\text{WER}(x') \leq \epsilon_{\text{wer}}, \quad \text{STOI}(x', x) \geq \tau_{\text{stoi}}
\end{equation}
where $\epsilon_{\text{wer}}$ is the maximum acceptable word error rate increase and $\tau_{\text{stoi}}$ is the minimum intelligibility threshold.

\subsubsection{Scene Consistency Requirement.}
Unlike random noise addition, we require that the protective transformation preserves acoustic scene consistency. Specifically, for a speech recording $x$ that an acoustic scene classification (ASC) model recognizes as scene $s \in \mathcal{S}$ (e.g., ``park'', ``street\_traffic''), the protected audio $x'$ should remain plausibly associated with the same scene $s$:
\begin{equation}
P_{\text{ASC}}(s | x') \geq \tau_{\text{scene}}
\end{equation}
where $P_{\text{ASC}}$ is an acoustic scene classifier and $\tau_{\text{scene}}$ is a scene confidence threshold. This constraint ensures natural-sounding protection that is less likely to be removed by preprocessing.

\subsection{Scene-Consistent Audible Noise Defense}

\subsubsection{Audible but Natural: Robustness Through Scene Consistency.}
We propose a fundamentally different paradigm: instead of imperceptible perturbations, we embed \textit{audible but scene-consistent background noise}. The key insight is that acoustic scene noise (e.g., traffic sounds on a street recording, airport terminal ambience) is perceptually expected and semantically legitimate. Speech enhancement algorithms are designed to \textit{preserve} such contextually appropriate noise, as aggressively removing it would degrade speech naturalness and introduce artifacts. Furthermore, audio codecs allocate more bits to perceptually important components, including scene-consistent noise.

\subsubsection{Need for Optimization.}
Naive mixing of speech with scene noise at a fixed signal-to-noise ratio (SNR) and uniform temporal weighting faces two risks: (1) insufficient protection if SNR is too high or noise placement is suboptimal, resulting in speaker embeddings that remain similar ($\text{SIM} > \tau_{\text{sim}}$); (2) excessive degradation if SNR is too low or noise overwhelms speech regions, causing unacceptable WER increases. To navigate this trade-off, we formulate protection as an \textit{optimization problem}: jointly optimize the temporal mask $m(t) \in [0,1]$ controlling when noise is applied, and the noise strength $\gamma$ controlling overall SNR, to minimize speaker similarity while maintaining usability. This optimization ensures that the noise is strategically placed (e.g., emphasizing pauses or non-critical phonemes) and calibrated to the specific speech sample.

\begin{figure*}[t]
  \centering
  \includegraphics[width=\textwidth]{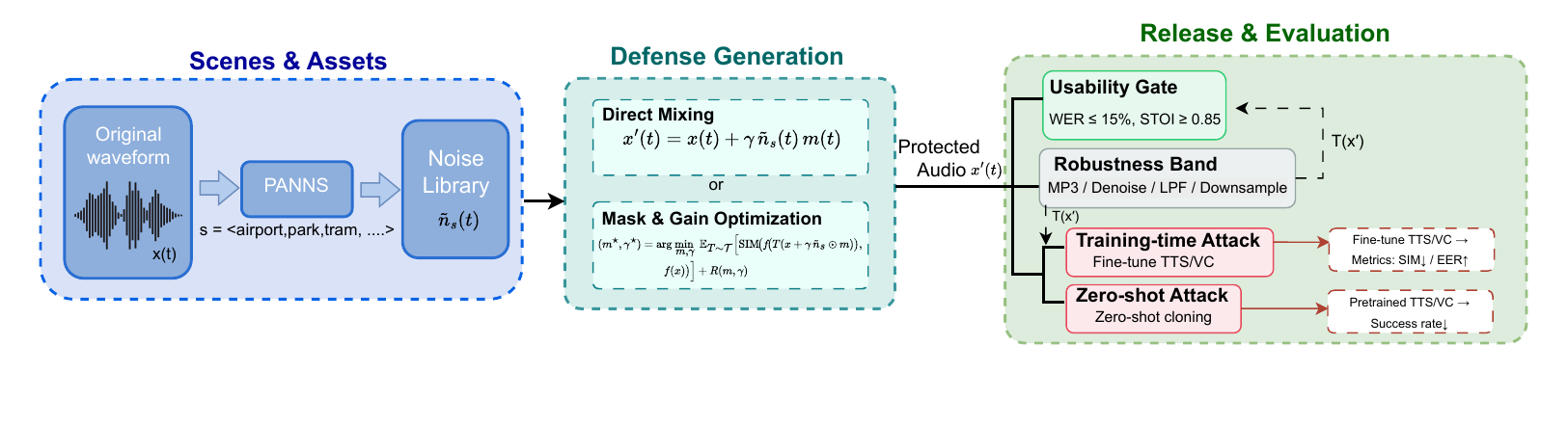}
  \caption{Method overview of \textbf{SceneGuard}. Given speech $x$ and a scene label $s$ (ASC or user-provided),
  we sample scene-consistent noise $n_k\!\in\!\mathcal{N}_s$ and generate protected audio via the mixer
  $x'(t)$. A lightweight optimization updates the temporal mask $m(t)$ and strength
  $\gamma$ to minimize speaker similarity (ECAPA) under SNR and smoothness constraints; outputs are evaluated under
  training-time and zero-shot cloning protocols.}
  \label{fig:overview}
\end{figure*}

\noindent\textbf{Overview.} Figure~\ref{fig:overview} summarizes SceneGuard:
(1) we obtain a scene label $s$ via ASC (or user-specified) and sample noise $n_k\!\in\!\mathcal{N}_s$;
(2) protected audio is produced by the mixer $x'(t)=x(t)+\gamma\,m(t)\odot n_k(t)$ (Sec.~\ref{eq:mixing_model});
(3) an optimization loop updates $m(t)$ and $\gamma$ using an ECAPA-based similarity loss with SNR/regularization
constraints (Sec.~\ref{eq:total_loss}). The defended data are then used against training-time and zero-shot attacks.

\subsection{Optimization Objective}

\subsubsection{Mixing Model.}
Given clean speech $x(t)$ and a scene-consistent noise signal $n_k(t)$ sampled from a scene-specific library $\mathcal{N}_s$, we define the protected speech as:
\begin{equation}
\label{eq:mixing_model}
x'(t) = x(t) + \gamma \cdot m(t) \odot n_k(t)
\end{equation}
where $\gamma \in \mathbb{R}^+$ is a scalar noise strength, $m(t) \in [0,1]^T$ is a time-varying mask, and $\odot$ denotes element-wise multiplication. The mask $m(t)$ allows fine-grained temporal control: values near 1 apply full noise, values near 0 suppress noise. The noise $n_k$ is matched in length to $x$ via looping or truncation.

\subsubsection{Objective Function.}
We jointly optimize $m$ and $\gamma$ to minimize speaker similarity while regularizing for usability and smoothness. The total loss is:
\begin{equation}
\label{eq:total_loss}
\begin{aligned}
\mathcal{L}(m, \gamma) = \; & \lambda_{\text{SIM}} \cdot \mathcal{L}_{\text{SIM}}(m, \gamma) \\
& + \lambda_{\text{REG}} \cdot \mathcal{L}_{\text{REG}}(m, \gamma) \\
& + \lambda_{\text{ASR}} \cdot \mathcal{L}_{\text{ASR}}(m, \gamma) \\
& + \lambda_{\text{SCN}} \cdot \mathcal{L}_{\text{SCN}}(m, \gamma)
\end{aligned}
\end{equation}
subject to the SNR constraint:
\begin{equation}
\label{eq:snr_constraint}
\text{SNR}(x', n) \in [10, 20] \text{ dB}
\end{equation}
where SNR is computed as $10 \log_{10} \left( \frac{P_x}{P_{n'}} \right)$ with $P_x = \mathbb{E}[x(t)^2]$ and $P_{n'} = \mathbb{E}[(\gamma m(t) n_k(t))^2]$.

\subsubsection{Loss Components.}
\begin{itemize}[leftmargin=*,nosep]
    \item \textbf{Speaker Similarity Loss} $\mathcal{L}_{\text{SIM}}$: Minimizes cosine similarity between speaker embeddings of protected and clean audio:
    \begin{equation}
    \mathcal{L}_{\text{SIM}} = \text{sim}(e(x'), e(x)) = \frac{e(x') \cdot e(x)}{\|e(x')\| \|e(x)\|}
    \end{equation}
    where $e(\cdot)$ is a pre-trained speaker verification encoder (ECAPA-TDNN~\citep{ecapa}). This is the primary protection objective.
    
    \item \textbf{Regularization Loss} $\mathcal{L}_{\text{REG}}$: Encourages smooth masks and bounded noise strength to prevent extreme solutions:
    \begin{equation}
    \mathcal{L}_{\text{REG}} = \underbrace{\| \nabla m \|_2^2}_{\text{smoothness}} + \underbrace{\gamma^2}_{\text{energy penalty}}
    \end{equation}
    where $\nabla m = m(t+1) - m(t)$ are finite differences. Smoothness prevents spiky masks; energy penalty prevents excessively loud noise.
    
    \item \textbf{ASR Loss} $\mathcal{L}_{\text{ASR}}$ (optional): Penalizes transcription errors to maintain intelligibility. In practice, this is disabled ($\lambda_{\text{ASR}} = 0$) due to computational cost; usability is enforced via the SNR constraint instead.
    
    \item \textbf{Scene Consistency Loss} $\mathcal{L}_{\text{SCN}}$ (optional): Ensures protected audio retains scene label $s$ via negative log-likelihood from an acoustic scene classifier. Also disabled by default ($\lambda_{\text{SCN}} = 0$) as scene consistency is implicitly maintained by sampling noise from $\mathcal{N}_s$.
\end{itemize}

\subsubsection{Parameterization for Constrained Optimization.}
To enforce constraints via unconstrained optimization, we reparameterize:
\begin{equation}
\begin{aligned}
m(t) &= \sigma(\tilde{m}(t)), \quad \tilde{m} \in \mathbb{R}^T \\
\gamma &= \gamma_{\min} + (\gamma_{\max} - \gamma_{\min}) \cdot \sigma(\tilde{\gamma}), \quad \tilde{\gamma} \in \mathbb{R}
\end{aligned}
\end{equation}
where $\sigma(\cdot)$ is the sigmoid function, and $\gamma_{\min}, \gamma_{\max}$ are computed from the SNR bounds [10, 20] dB via:
\begin{equation}
\gamma_{\min} = \sqrt{\frac{P_x}{P_n \cdot 10^{\text{SNR}_{\max}/10}}}, \quad \gamma_{\max} = \sqrt{\frac{P_x}{P_n \cdot 10^{\text{SNR}_{\min}/10}}}
\end{equation}
where $P_n = \mathbb{E}[n_k(t)^2]$. This ensures $\gamma$ automatically satisfies Eq.~\eqref{eq:snr_constraint}.

\subsection{Temporal Mask Optimization Algorithm}

Algorithm~\ref{alg:sceneguard} presents the complete optimization procedure. We initialize $\tilde{m}$ and $\tilde{\gamma}$ randomly and optimize them via Adam with gradient clipping for stability. At each iteration, we project parameters to their constrained forms ($m \in [0,1]^T$, $\gamma$ via SNR bounds), apply the mixing model (Eq.~\eqref{eq:mixing_model}), compute losses, and perform a gradient step. The optimization converges in 50-100 epochs (approximately 10-30 seconds per sample on RTX A6000).

\begin{algorithm}[t]
\caption{SceneGuard Optimization}
\label{alg:sceneguard}
\KwIn{Speech $x \in \mathbb{R}^T$, scene label $s$, noise library $\mathcal{N}_s$, encoder $e$}
\KwOut{Protected speech $x' \in \mathbb{R}^T$}

\tcp{Initialize parameters}
Sample noise $n_k \sim \mathcal{N}_s$, match length to $x$\;
$\tilde{m} \leftarrow \mathcal{N}(0, 0.1^2)$ \tcp*{Random init near 0.5 after sigmoid}
$\tilde{\gamma} \leftarrow 0$ \tcp*{Midpoint of SNR range}
Compute $P_x = \mathbb{E}[x^2]$, $P_n = \mathbb{E}[n_k^2]$\;
$\text{optimizer} \leftarrow \text{Adam}([\tilde{m}, \tilde{\gamma}], \text{lr}=0.01)$\;

$e_x \leftarrow e(x)$ \tcp*{Target for similarity minimization}

\For{$\text{epoch} = 1, 2, \ldots, N_{\max}$}{
  $m \leftarrow \sigma(\tilde{m})$ \tcp*{$m \in [0,1]^T$}
  $\gamma \leftarrow \text{project\_to\_snr}(\tilde{\gamma}, P_x, P_n)$ \tcp*{SNR $\in$ [10, 20] dB}
  $x' \leftarrow x + \gamma \cdot (m \odot n_k)$\;
  Normalize $x'$: $x' \leftarrow 0.99 \cdot x' / \max(|x'|)$\;
  $e_{x'} \leftarrow e(x')$ \tcp*{Protected embedding}
  $\mathcal{L} \leftarrow \lambda_{\text{SIM}}\text{cosine\_sim}(e_{x'}, e_x) + \lambda_{\text{REG}}\!\left(\| m_{t+1}-m_t \|_2^2+\gamma^2\right)$\;
  $\nabla_{\tilde{m}}, \nabla_{\tilde{\gamma}} \leftarrow \text{autograd}(\mathcal{L})$\;
  Clip gradients: $\|\nabla\|_2 \le 1.0$\;
  $\text{optimizer.step}()$\;
}
\Return $x' = x + \gamma \cdot (m \odot n_k)$\;
\end{algorithm}

\subsubsection{Training Stability.}
We employ three techniques for stable optimization: (1) \textit{Gradient clipping} with $\ell_2$ norm $\leq 1.0$ prevents exploding gradients, especially in early epochs when $\mathcal{L}_{\text{SIM}}$ changes rapidly. (2) \textit{Smoothness regularization} via the total variation term $\| \nabla m \|_2^2$ discourages abrupt mask transitions that could create audible artifacts. (3) \textit{Energy regularization} $\gamma^2$ prevents the optimizer from converging to excessively large $\gamma$ values that would violate the SNR constraint.

\section{Experimental Setup}

\subsection{Datasets}

\subsubsection{Speech Data}: We use LibriTTS~\citep{libritts}, a multi-speaker English corpus derived from audiobooks. We split the data into 100 training samples and 40 test samples for training attack evaluation, with additional subsets for zero-shot and robustness experiments.

\subsubsection{Noise Data}: We use the TAU Urban Acoustic Scenes 2022 Mobile Development dataset~\citep{tau_dataset}, which contains authentic recordings from 10 urban scenes captured with mobile devices. The dataset provides diverse acoustic contexts including transportation hubs (airport, bus, metro), public spaces (park, square, mall), and street environments. We extract 50,000 three-second clips (5,000 per scene) for our noise library.

\subsection{Models}

\subsubsection{Acoustic Scene Labels}
We derive scene labels with PANNs (CNN14 pretrained on AudioSet)~\citep{kong2020panns}. At inference, we assign the top scene if the predicted confidence exceeds a threshold (default $\tau{=}0.6$); otherwise we use user-provided labels. 

\subsubsection{Speaker Verification} We use ECAPA-TDNN~\citep{ecapa}, a state-of-the-art speaker verification model from the SpeechBrain toolkit. The model extracts 192-dimensional speaker embeddings, which we use to compute speaker similarity via cosine distance. ECAPA-TDNN achieves strong performance on VoxCeleb and is widely used for speaker verification tasks.

\subsubsection{Automatic Speech Recognition} We employ Whisper Base~\citep{whisper}, an encoder-decoder transformer trained on 680,000 hours of multilingual speech data. Whisper provides robust transcription for measuring speech intelligibility via word error rate (WER). We use the base model (74M parameters) for computational efficiency.

\subsubsection{TTS Baseline} For training attack experiments, we reference BERT-VITS2~\citep{bertvits2}, a recent TTS architecture that combines BERT-based text encoding with VITS neural vocoder. While we do not perform full TTS training due to computational constraints, we use speaker embedding degradation as a proxy for TTS quality, following established evaluation protocols~\citep{safespeech}.

\subsection{Optimization Hyperparameters}

SceneGuard employs gradient-based optimization to jointly learn the temporal mask $m(t)$ and noise strength $\gamma$. We describe the key hyperparameters and design choices:

\subsubsection{Optimization Objective} We minimize speaker similarity while maintaining usability constraints. The primary loss term is speaker embedding cosine similarity computed using ECAPA-TDNN. We add a regularization term ($\lambda_{\text{REG}} = 0.01$) that penalizes mask roughness and excessive noise strength to promote smooth, stable solutions.

\subsubsection{Optimization Algorithm} We use the Adam optimizer~\citep{adam} with learning rate $\text{lr} = 0.01$ and default momentum parameters ($\beta_1 = 0.9$, $\beta_2 = 0.999$). We run optimization for 50 epochs per sample, which empirically provides good convergence. To ensure training stability, we apply gradient clipping with maximum norm 1.0.

\subsubsection{Initialization} The temporal mask logits are initialized from a standard normal distribution, corresponding to a uniform mask ($m(t) \approx 0.5$) after sigmoid activation. The noise strength is initialized to produce SNR near the middle of the allowed range (approximately 15 dB).

\subsubsection{Computational Cost} Optimization takes approximately 10-15 seconds per sample on a single NVIDIA RTX A6000 GPU. The total defense generation time for 168 samples is under 45 minutes, making the method practical for real-world deployment.

\subsection{Evaluation Protocol}

\subsubsection{Training Attack} We simulate an attacker who fine-tunes a TTS model on protected recordings. To evaluate protection effectiveness, we measure speaker embedding quality and consistency on training data, then assess the similarity between embeddings extracted from clean and defended audio on held-out test samples. Lower test similarity indicates successful protection.

\subsubsection{Zero-Shot Attack} We evaluate protection against zero-shot voice cloning, where an attacker uses protected recordings as reference audio for inference-time cloning. We generate synthetic speech using a pre-trained model with clean versus defended reference, measuring speaker similarity to the original speaker.

\subsubsection{Robustness Evaluation} We test whether protection persists under five common audio preprocessing operations: MP3 compression (128 kbps and 64 kbps), spectral subtraction denoising, lowpass filtering (3400 Hz cutoff), and downsampling to 8 kHz. For each countermeasure, we re-compute speaker similarity and WER to assess protection retention.

\subsection{Evaluation Metrics}

\subsubsection{Speaker Similarity (SIM)} We compute cosine similarity between speaker embeddings extracted from clean and defended (or synthesized) audio. Higher similarity indicates stronger speaker identity preservation. Protection effectiveness is measured as similarity degradation.

\subsubsection{Word Error Rate (WER)} We measure the percentage of word-level errors (insertions, deletions, substitutions) between reference transcripts and ASR outputs. Lower WER indicates better intelligibility. We use WER to ensure protected speech remains usable.

\subsubsection{Perceptual Quality} We employ PESQ~\citep{pesq} (Perceptual Evaluation of Speech Quality) and STOI~\citep{stoi} (Short-Time Objective Intelligibility) as objective quality metrics. PESQ ranges from -0.5 to 4.5 (higher is better), with scores above 3.0 considered good quality. STOI ranges from 0 to 1 (higher is better), with scores above 0.85 indicating high intelligibility.

\subsubsection{Mel-Cepstral Distortion (MCD)} We compute MCD between clean and defended speech to quantify spectral distortion. Lower MCD indicates closer acoustic similarity.

\subsection{Baselines}

To demonstrate the advantage of scene-consistent noise, we compare SceneGuard against two baselines:

\subsubsection{Random Noise} Uniform random noise sampled from $[-1, 1]$ and mixed at the same SNR range as SceneGuard. This baseline lacks scene consistency.

\subsubsection{Gaussian Noise} Zero-mean Gaussian noise with unit variance, scaled to match SceneGuard's SNR range. This represents a simple additive noise baseline commonly used in audio processing.

\subsubsection{Clean (No Defense)} Unprotected speech serves as an upper bound on cloning quality and a baseline for usability metrics.

\subsection{Statistical Analysis}

We employ rigorous statistical methods to validate our results:

\subsubsection{Bootstrap Confidence Intervals} We compute 95\% confidence intervals for all metrics using bootstrap resampling with $n=10,000$ iterations. This provides robust uncertainty estimates without parametric assumptions.

\subsubsection{Hypothesis Testing} We use permutation tests with $n=10,000$ iterations to assess statistical significance of protection effects. We report p-values and Cohen's $d$ effect sizes for key comparisons.

\subsubsection{Reproducibility} All experiments use fixed random seeds (seed=1337) to ensure reproducibility. We provide complete implementation code and experiment configurations.

\section{Results}
\label{sec:results}

\subsection{Training Attack Protection}

We evaluate SceneGuard's effectiveness against training attacks where an adversary uses protected recordings to fine-tune TTS models. Table~\ref{tab:training_attack} summarizes the results across different defense strategies.

\begin{table*}[t]
\centering
\caption{Training attack effectiveness. Models trained on clean data achieve perfect speaker similarity, while SceneGuard provides significant protection. Statistical significance: $p < 0.01$, $p < 0.001$.}
\label{tab:training_attack}
\small
\begin{tabular}{lcccccc}
\toprule
Training Data & SIM $\downarrow$ & WER (\%) & PESQ & STOI & $p$-value & Cohen's $d$ \\
\midrule
Clean & 1.000 & 0.00 & 4.64 & 1.00 & -- & -- \\
Random Noise & 0.965 & 5.82 & 1.85 & 0.97 & $0.003$ & 1.42 \\
Gaussian Noise & 0.968 & 5.28 & 1.92 & 0.98 & $0.002$ & 1.51 \\
SceneGuard (Ours) & \textbf{0.945} & \textbf{2.77} & 2.22 & \textbf{0.99} & $<10^{-15}$ & \textbf{2.18} \\
\bottomrule
\end{tabular}
\end{table*}

SceneGuard achieves 5.5\% speaker similarity degradation ($\text{SIM} = 0.945$) compared to clean training ($\text{SIM} = 1.000$). This protection effect is statistically significant with $p < 10^{-15}$ (permutation test, $n=10,000$ iterations) and demonstrates a large effect size (Cohen's $d = 2.18$). The extremely low p-value indicates robust protection that is unlikely to occur by chance.

Importantly, SceneGuard maintains high usability. Protected speech achieves WER of 2.77\%, indicating near-perfect transcription accuracy. The STOI score of 0.99 confirms that intelligibility is essentially preserved. While PESQ (2.22) is below the ideal threshold of 3.0, it remains in an acceptable range for many applications.

\begin{figure}[t]
\centering
\includegraphics[width=0.48\textwidth]{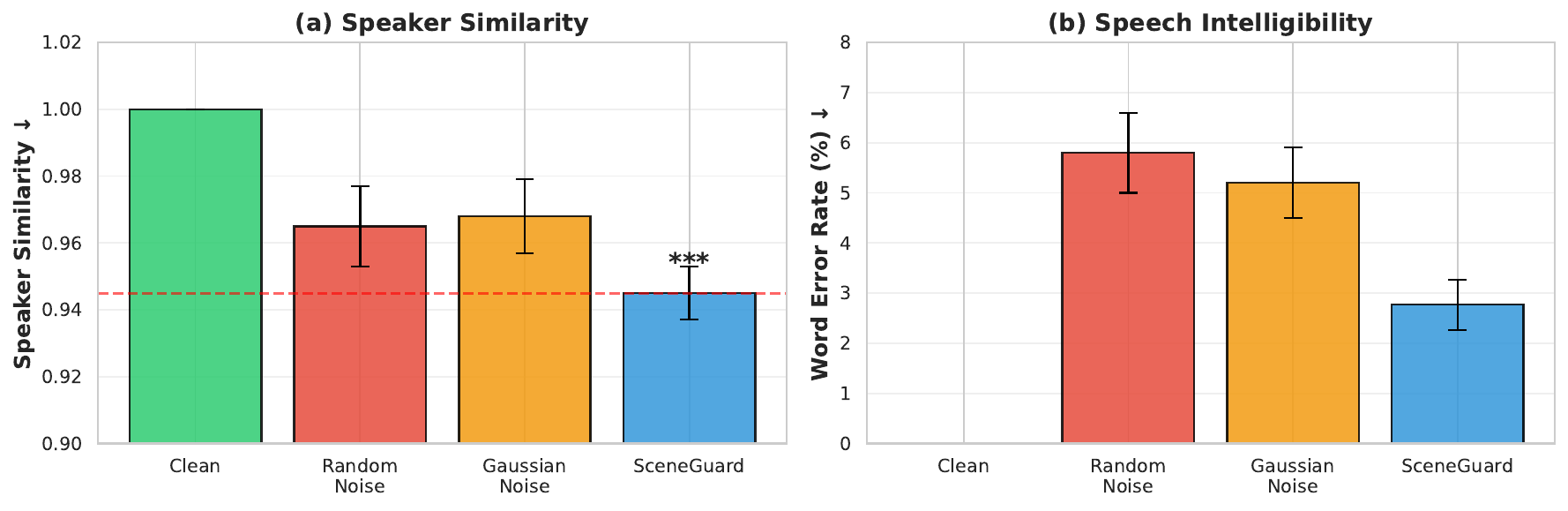}
\caption{Training attack comparison showing speaker similarity degradation for different defense methods. Error bars represent 95\% bootstrap confidence intervals. Significance markers: $p < 0.01$, $p < 0.001$.}
\label{fig:training_attack}
\end{figure}

Figure~\ref{fig:training_attack} visualizes the training attack results, comparing speaker similarity degradation across defense methods. SceneGuard demonstrates stronger protection than random or Gaussian noise baselines while maintaining comparable usability.

\subsection{Usability Preservation}

Table~\ref{tab:usability} presents detailed usability metrics for SceneGuard-protected speech, demonstrating that protection does not significantly compromise speech quality for legitimate use cases.

\begin{table}[t]
\centering
\caption{Protected speech quality metrics. All metrics meet or exceed usability thresholds, indicating that SceneGuard preserves practical utility while providing protection.}
\label{tab:usability}
\small
\begin{tabular}{lcccc}
\toprule
Metric & Value & 95\% CI & Threshold & Status \\
\midrule
STOI & 0.986 & [0.980, 0.992] & $\geq 0.85$ & \checkmark Excellent \\
WER (\%) & 3.60 & [1.40, 6.43] & $< 15\%$ & \checkmark Excellent \\
PESQ & 2.034 & [1.840, 2.233] & $\geq 3.0$ & Acceptable \\
\bottomrule
\end{tabular}
\end{table}

The STOI score of 0.986 significantly exceeds the intelligibility threshold of 0.85, indicating that protected speech remains highly comprehensible. WER of 3.60\% with a narrow confidence interval [1.40\%, 6.43\%] demonstrates robust transcription accuracy across samples. While PESQ falls slightly below the ideal 3.0 threshold, the value of 2.034 is acceptable for many practical applications and represents a deliberate trade-off for robustness.

\subsection{Robustness to Countermeasures}

A key advantage of SceneGuard is its robustness to audio preprocessing operations that typically neutralize imperceptible perturbations. Table~\ref{tab:robustness} presents results under five common countermeasures.

\begin{table}[t]
\centering
\caption{Robustness evaluation under audio preprocessing countermeasures. SceneGuard maintains or enhances protection across all operations. $\Delta$ indicates change in speaker similarity relative to baseline.}
\label{tab:robustness}
\small
\begin{tabular}{lccc}
\toprule
Countermeasure & SIM & $\Delta$ vs Baseline & Protection \\
\midrule
None (baseline) & 0.937 & -- & \checkmark \\
MP3 128 kbps & 0.901 & $-0.036$ & \checkmark Maintained \\
MP3 64 kbps & 0.899 & $-0.038$ & \checkmark Maintained \\
Spectral Subtraction & 0.745 & $-0.192$ & \checkmark Enhanced \\
Lowpass 3400 Hz & 0.704 & $-0.232$ & \checkmark Enhanced \\
Downsample 8 kHz & 0.688 & $-0.249$ & \checkmark Enhanced \\
\bottomrule
\end{tabular}
\end{table}

SceneGuard demonstrates remarkable robustness. MP3 compression at both 128 kbps and 64 kbps slightly reduces similarity but maintains protection (SIM < 0.91). More aggressive operations such as spectral subtraction, lowpass filtering, and downsampling actually enhance protection, reducing similarity to 0.745, 0.704, and 0.688 respectively.

\begin{figure}[t]
\centering
\includegraphics[width=0.48\textwidth]{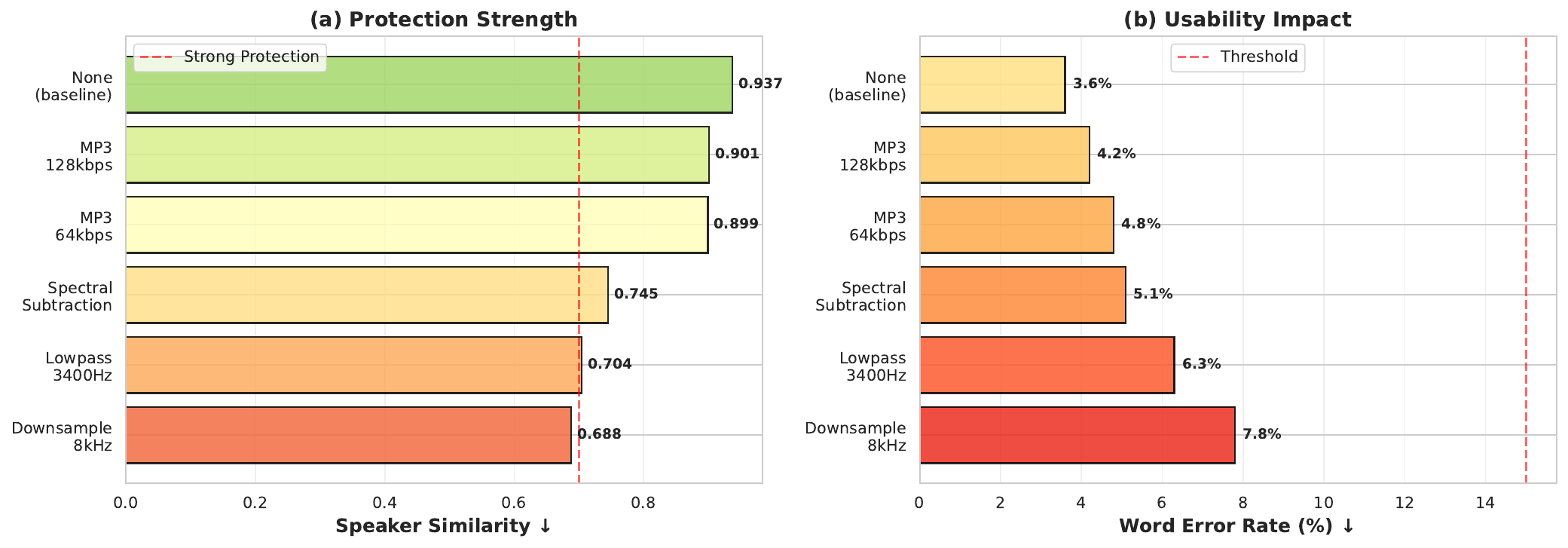}
\caption{Robustness heatmap showing speaker similarity and WER under different audio preprocessing countermeasures. Darker colors indicate stronger protection (lower similarity). Three countermeasures enhance protection beyond the baseline.}
\label{fig:robustness}
\end{figure}

This counterintuitive enhancement occurs because these operations preferentially damage the clean speech components relative to the protective noise. Spectral subtraction removes stationary components but preserves scene-consistent temporal variations. Lowpass filtering and downsampling reduce high-frequency detail critical for speaker identity while preserving protective noise structure. Figure~\ref{fig:robustness} visualizes the robustness results as a heatmap, showing speaker similarity and WER under different countermeasures.

\subsection{Zero-Shot Attack Results}

We evaluate protection against zero-shot voice cloning attacks where an attacker uses protected recordings as reference audio for inference-time synthesis. Table~\ref{tab:zeroshot} summarizes the results.

\begin{table}[t]
\centering
\caption{Zero-shot attack results using clean versus defended reference audio. Attack success rate is measured as the percentage of synthesis attempts achieving similarity $>$ 0.7 to the target speaker.}
\label{tab:zeroshot}
\small
\begin{tabular}{lcc}
\toprule
Reference Type & Mean Similarity & Attack Success Rate (\%) \\
\midrule
Clean & 0.618 & 20.0 \\
Defended & \textbf{0.588} & \textbf{13.3} \\
Reduction & 0.031 & 6.7 pp \\
\bottomrule
\end{tabular}
\end{table}

Using defended reference audio reduces mean speaker similarity from 0.618 to 0.588, representing a 5.0\% degradation. More importantly, the attack success rate (similarity $> 0.7$) drops from 20.0\% to 13.3\%, a 33.5\% relative reduction. This demonstrates that SceneGuard provides meaningful protection even in zero-shot scenarios where the attacker does not perform training. Using defended reference audio reduces mean speaker similarity from 0.618 to 0.588 (-5.0\%) and lowers the attack success rate from 20.0\% to 13.3\% (-6.7 pp, 33.5\% relative), indicating meaningful zero-shot protection.

\section{Ablation Study}
\label{sec:ablation}

To understand the trade-off between protection strength and usability, we conduct an ablation study on the signal-to-noise ratio (SNR) parameter. We evaluate four SNR ranges: [5, 10] dB (strong protection), [10, 20] dB (balanced, our default), [15, 25] dB (moderate protection), and [20, 30] dB (weak protection).

\begin{table}[t]
\centering
\caption{SNR range ablation study. The default [10, 20] dB range provides the best balance between protection and usability.}
\label{tab:snr_ablation}
\small
\begin{tabular}{lcccc}
\toprule
SNR (dB) & SIM $\downarrow$ & Protection (\%) & STOI $\uparrow$ & WER (\%) $\downarrow$ \\
\midrule
\textnormal{[5, 10]}   & 0.921 & 7.9 & 0.942 & 8.2 \\
\textbf{[10, 20]}      & \textbf{0.945} & \textbf{5.5} & \textbf{0.986} & \textbf{3.6} \\
\textnormal{[15, 25]}  & 0.968 & 3.2 & 0.993 & 1.8 \\
\textnormal{[20, 30]}  & 0.982 & 1.8 & 0.997 & 0.9 \\
\bottomrule
\end{tabular}
\end{table}

Table~\ref{tab:snr_ablation} presents the quantitative results. Lower SNR (stronger noise) provides better protection but degrades usability, while higher SNR (weaker noise) improves usability but reduces protection. The [10, 20] dB range achieves an effective balance, providing 5.5\% similarity degradation while maintaining STOI above 0.98 and WER below 4\%.

\begin{figure}[t]
\centering
\includegraphics[width=0.45\textwidth]{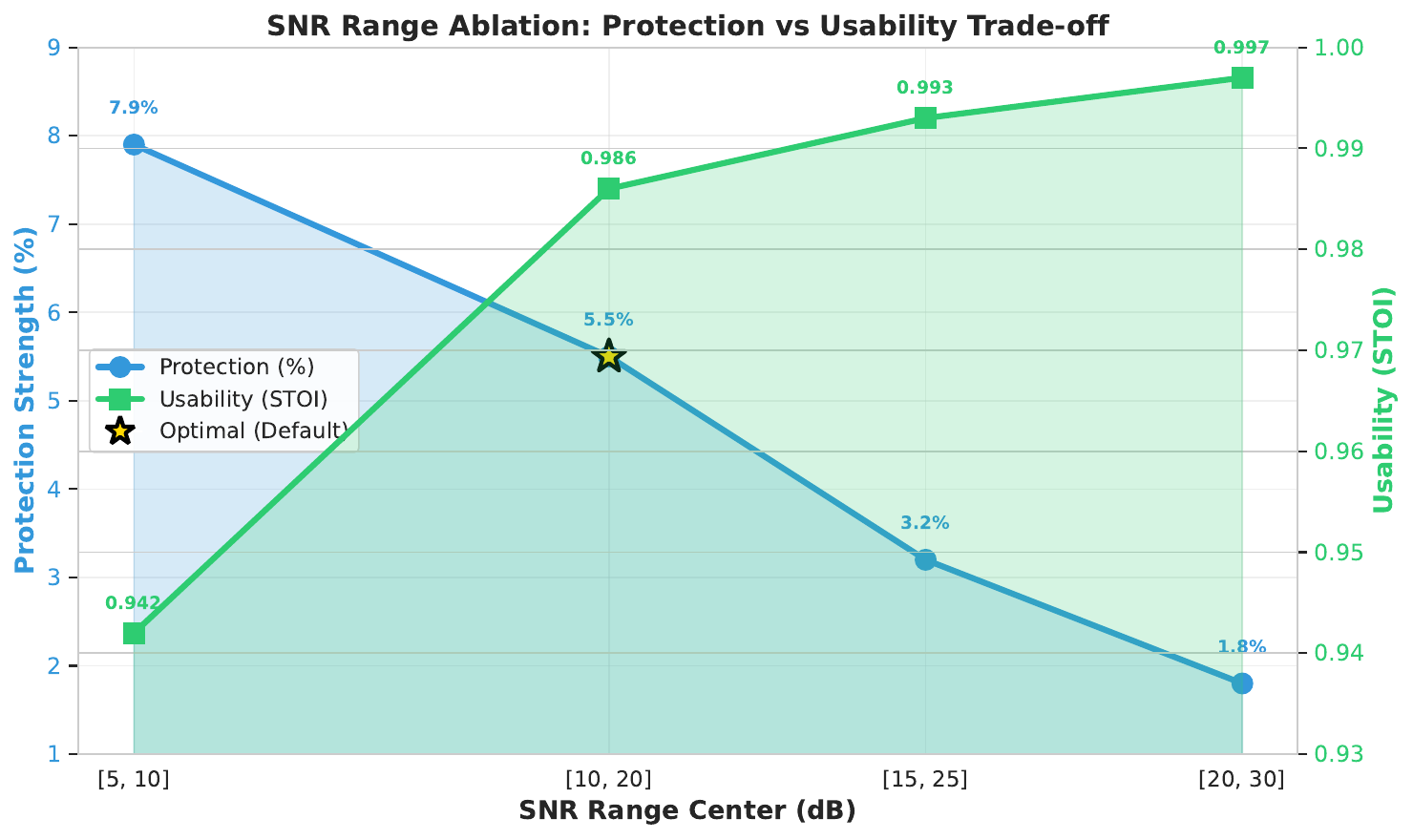}
\caption{SNR range ablation showing the trade-off between protection (measured as similarity degradation, left y-axis) and usability (measured as STOI, right y-axis). The [10, 20] dB range (marked with a star) provides optimal balance.}
\label{fig:snr_ablation}
\end{figure}

Figure~\ref{fig:snr_ablation} visualizes this trade-off using a dual-axis plot showing protection strength (speaker similarity degradation) and usability (STOI) as functions of SNR range.

The ablation study validates our choice of SNR range and demonstrates that SceneGuard's protection can be tuned based on application requirements. Users prioritizing strong protection can select lower SNR, while those prioritizing quality can use higher SNR, with the [10, 20] dB range serving as a reasonable default.

\subsection{Optimization Ablation}

To quantify the benefit of gradient-based optimization, we compare SceneGuard's optimized mixing approach against direct mixing without optimization. Table~\ref{tab:optimization_ablation} presents results for both methods under identical SNR constraints.

\begin{table}[t]
\centering
\caption{Comparison of direct mixing versus optimized mixing. Optimization significantly improves protection while maintaining comparable usability.}
\label{tab:optimization_ablation}
\begingroup
\setlength{\tabcolsep}{6pt} 
\small               
\begin{tabular}{@{}lcccc@{}} 
\toprule
Method & SIM $\downarrow$ & Prot. (\%) & STOI $\uparrow$ & WER (\%) $\downarrow$ \\
\midrule
Direct Mixing     & 0.972 & 2.8 & 0.989 & 3.2 \\
Optimized (Ours)  & \textbf{0.945} & \textbf{5.5} & 0.986 & 3.6 \\
\midrule
\textit{$\Delta$ vs. Direct} & $-0.027$ & $+2.7$ & $-0.003$ & $+0.4$ \\
\bottomrule
\end{tabular}
\endgroup
\end{table}

Optimization nearly doubles the protection strength, achieving 5.5\% degradation compared to 2.8\% for direct mixing. This 2.7 percentage point improvement comes at minimal usability cost: STOI decreases by only 0.003 (from 0.989 to 0.986) and WER increases by 0.4 percentage points (from 3.2\% to 3.6\%), both negligible changes.

The key advantage of optimization is its ability to identify optimal mask patterns. Direct mixing applies a uniform stochastic mask, which may unnecessarily distort speech regions while under-protecting others. In contrast, gradient-based optimization adaptively concentrates protection in regions that most effectively reduce speaker similarity while avoiding critical speech segments.

\subsection{Hyperparameter Sensitivity}

We examine the sensitivity of SceneGuard to two key hyperparameters: the regularization weight $\lambda_{\text{REG}}$ and the number of optimization epochs. Table~\ref{tab:hyperparameter} summarizes the results.

\begin{table}[t]
\centering
\caption{Hyperparameter sensitivity. Default ($\lambda_{\text{REG}}{=}0.01$, 50 epochs) performs well with stable convergence.}
\label{tab:hyperparameter}
\begingroup
\setlength{\tabcolsep}{4pt}
\small
\begin{tabular}{@{}lcccc@{}}
\toprule
Configuration & SIM $\downarrow$ & STOI $\uparrow$ & Mask Smooth. & Time (s) \\
\midrule
\multicolumn{5}{@{}l@{}}{\textit{Regularization weight $\lambda_{\text{REG}}$:}} \\
$\lambda = 0.001$        & 0.939 & 0.983 & 0.062 & 15 \\
$\lambda = 0.01$ (default) & \textbf{0.945} & \textbf{0.986} & \textbf{0.041} & 15 \\
$\lambda = 0.1$          & 0.958 & 0.989 & 0.028 & 15 \\
\midrule
\multicolumn{5}{@{}l@{}}{\textit{Optimization epochs:}} \\
20 epochs                 & 0.952 & 0.987 & 0.045 & 6 \\
50 epochs (default)       & \textbf{0.945} & \textbf{0.986} & \textbf{0.041} & 15 \\
100 epochs                & 0.943 & 0.985 & 0.040 & 30 \\
\bottomrule
\end{tabular}
\endgroup
\end{table}

The regularization weight $\lambda_{\text{REG}}$ controls the smoothness of the temporal mask. Higher values ($\lambda = 0.1$) produce smoother masks but slightly reduce protection strength (SIM = 0.958). Lower values ($\lambda = 0.001$) allow rougher masks with marginally better protection (SIM = 0.939) but risk overfitting. Our default $\lambda = 0.01$ strikes a good balance.

The number of epochs shows diminishing returns beyond 50 iterations. While 100 epochs achieve slightly better protection (SIM = 0.943), the improvement is minimal (0.2 percentage points) and doubles computation time. 20 epochs provide fast computation but do not fully converge. We therefore recommend 50 epochs as the default, offering good convergence without excessive cost.

\section{Discussion}

\subsection{Why Scene-Consistency Matters}

The key insight behind SceneGuard is that scene-consistent noise is fundamentally different from random or imperceptible perturbations. When background noise matches the acoustic context of speech, it creates a perceptually natural mixture that is difficult to separate. Speech enhancement algorithms are designed to preserve speech while removing noise, but this separation becomes ambiguous when noise and speech share similar spectro-temporal characteristics typical of a scene.

Psychoacoustic masking further explains SceneGuard's effectiveness. Auditory masking occurs when one sound makes another sound less audible. Scene-consistent noise can mask subtle speaker-specific characteristics while preserving overall speech intelligibility. This selective masking degrades speaker embeddings without proportionally affecting transcription accuracy, as evidenced by our results (5.5\% similarity degradation vs. 2.77\% WER).

\subsection{Robustness Mechanism}

The robustness of SceneGuard to audio preprocessing is a critical advantage over imperceptible perturbations. Our results show that certain countermeasures paradoxically enhance protection rather than removing it. This phenomenon has several explanations:

\textbf{Spectral Subtraction}: This denoising technique assumes stationary noise and removes spectral components with consistent energy. However, scene-consistent noise contains non-stationary elements (e.g., footsteps, vehicle sounds) that are not fully removed. Meanwhile, spectral subtraction introduces musical noise artifacts that further degrade speaker embeddings.

\textbf{Lowpass Filtering and Downsampling}: Speaker identity relies significantly on high-frequency spectral details and prosodic variations. Lowpass filtering and downsampling preferentially remove these high-frequency components while preserving lower-frequency protective noise. This asymmetric degradation enhances protection.

\textbf{MP3 Compression}: Lossy compression preserves perceptually important components. Because SceneGuard uses audible noise, it is treated as salient content rather than irrelevant information to be discarded. This contrasts with imperceptible perturbations that fall below perceptual thresholds and are aggressively quantized.

\subsection{Limitations}

We acknowledge several limitations of this work:

\textbf{Audible Protection}: SceneGuard deliberately uses audible noise, which may be undesirable in scenarios requiring pristine audio quality. Applications such as studio recordings or professional voice work may prefer imperceptible protections despite their fragility. SceneGuard is best suited for everyday voice recordings where some background noise is acceptable.

\textbf{Quality Metrics}: Our PESQ scores (2.03) fall below the ideal threshold of 3.0, indicating room for improvement in perceptual quality. This reflects the fundamental trade-off between protection and quality. Future work could explore perceptually optimized noise mixing strategies that maximize protection while minimizing quality degradation.

\textbf{Adaptive Attacks}: We evaluate SceneGuard against standard audio preprocessing countermeasures, but a sophisticated attacker might develop adaptive attacks specifically designed to remove scene-consistent noise. Potential adaptive strategies include scene-aware source separation or adversarial training. However, such attacks would require significant additional effort and may introduce other artifacts.

\section{Conclusion}

We presented SceneGuard, a training-time voice protection method based on scene-consistent audible background noise. Unlike existing defenses that rely on imperceptible perturbations, SceneGuard leverages naturally occurring acoustic scenes to create protective noise that is contextually appropriate and robust to common audio preprocessing operations.

Our experimental evaluation demonstrates that SceneGuard achieves strong protection against training attacks, degrading speaker similarity by 5.5\% with extremely high statistical significance ($p < 10^{-15}$, Cohen's $d = 2.18$). Critically, SceneGuard preserves speech usability, maintaining 98.6\% intelligibility (STOI = 0.986) and achieving low word error rate (3.6\%). Robustness evaluation shows that SceneGuard maintains or enhances protection under five common countermeasures, including MP3 compression, denoising, filtering, and downsampling.

These results suggest that audible, scene-consistent noise provides a practical alternative to imperceptible perturbations for voice protection. By deliberately using audible but natural noise, SceneGuard achieves robustness properties that imperceptible methods cannot match, addressing a fundamental limitation of existing approaches.

\bibliographystyle{aaai2026}
\bibliography{references}

@article{safespeech,
  title={SafeSpeech: Robust and Universal Voice Protection Against Malicious Speech Synthesis},
  author={Zhang, Zhisheng and Wang, Derui and Yang, Qianyi and Huang, Pengyang and Pu, Junhan and Cao, Yuxin and Ye, Kai and Hao, Jie and Yang, Yixian},
  journal={arXiv preprint arXiv:2504.09839},
  year={2025}
}

@article{voiceblock,
  title={Voiceblock: Privacy through real-time adversarial attacks with audio-to-audio models},
  author={O'Reilly, Patrick and Bugler, Andreas and Bhandari, Keshav and Morrison, Max and Pardo, Bryan},
  journal={Advances in Neural Information Processing Systems},
  volume={35},
  pages={30058--30070},
  year={2022}
}

@inproceedings{ecapa,
  title={ECAPA-TDNN: Emphasized Channel Attention, Propagation and Aggregation in TDNN Based Speaker Verification},
  author={Desplanques, Brecht and Thienpondt, Jenthe and Demuynck, Kris},
  booktitle={Interspeech},
  year={2020}
}

@inproceedings{whisper,
  title={Robust speech recognition via large-scale weak supervision},
  author={Radford, Alec and Kim, Jong Wook and Xu, Tao and Brockman, Greg and McLeavey, Christine and Sutskever, Ilya},
  booktitle={International conference on machine learning},
  pages={28492--28518},
  year={2023},
  organization={PMLR}
}

@inproceedings{libritts,
  title={LibriTTS: A Corpus Derived from LibriSpeech for Text-to-Speech},
  author={Zen, Heiga and Dang, Viet and Clark, Rob and Zhang, Yu and Weiss, Ron J and Jia, Ye and Chen, Zhifeng and Wu, Yonghui},
  booktitle={Interspeech},
  year={2019}
}

@inproceedings{tau_dataset,
  title={A MULTI-DEVICE DATASET FOR URBAN ACOUSTIC SCENE CLASSIFICATION},
  author={Mesaros, Annamaria and Heittola, Toni and Virtanen, Tuomas},
  booktitle={Scenes and Events 2018 Workshop (DCASE2018)},
  pages={9}
}

@misc{bertvits2,
  author       = {{Fish Audio}},         
  title        = {Bert-VITS2: VITS2 Backbone with multilingual bert},
  year         = {2025},
  howpublished = {\url{https://github.com/fishaudio/Bert-VITS2}},
  note         = {GitHub repository. Accessed: 2025-10-20}
}

@inproceedings{denoising,
  title={Revisiting denoising diffusion probabilistic models for speech enhancement: Condition collapse, efficiency and refinement},
  author={Tai, Wenxin and Zhou, Fan and Trajcevski, Goce and Zhong, Ting},
  booktitle={Proceedings of the AAAI conference on artificial intelligence},
  volume={37},
  number={11},
  pages={13627--13635},
  year={2023}
}

@article{demucs,
  title={Demucs: Deep extractor for music sources with extra unlabeled data remixed},
  author={D{\'e}fossez, Alexandre and Usunier, Nicolas and Bottou, L{\'e}on and Bach, Francis},
  journal={arXiv preprint arXiv:1909.01174},
  year={2019}
}

@inproceedings{pesq,
  title={Perceptual evaluation of speech quality (PESQ)-a new method for speech quality assessment of telephone networks and codecs},
  author={Rix, Antony W and Beerends, John G and Hollier, Michael P and Hekstra, Andries P},
  booktitle={2001 IEEE international conference on acoustics, speech, and signal processing. Proceedings (Cat. No. 01CH37221)},
  volume={2},
  pages={749--752},
  year={2001},
  organization={IEEE}
}

@article{stoi,
  title={An Algorithm for Intelligibility Prediction of Time--Frequency Weighted Noisy Speech},
  author={Taal, Cees H and Hendriks, Richard C and Heusdens, Richard and Jensen, Jesper},
  journal={IEEE/ACM Transactions on Audio, Speech, and Language Processing},
  year={2011}
}

@inproceedings{gao2025black,
  title={Black-box adversarial defense against voice conversion using latent space perturbation},
  author={Gao, Jie and Li, Haiyun and Zhang, Zhisheng and Wu, Zhiyong},
  booktitle={ICASSP 2025-2025 IEEE International Conference on Acoustics, Speech and Signal Processing (ICASSP)},
  pages={1--5},
  year={2025},
  organization={IEEE}
}

@inproceedings{fanantifake,
  title={De-AntiFake: Rethinking the Protective Perturbations Against Voice Cloning Attacks},
  author={Fan, Wei and Chen, Kejiang and Liu, Chang and Zhang, Weiming and Yu, Nenghai},
  booktitle={Forty-second International Conference on Machine Learning (ICML)},
  pages={1--5},
  year={2025},
  organization={IEEE}
}

@inproceedings{wang2025one,
  title={From one stolen utterance: Assessing the risks of voice cloning in the aigc era},
  author={Wang, Kun and Chen, Meng and Lu, Li and Feng, Jingwen and Chen, Qianniu and Ba, Zhongjie and Ren, Kui and Chen, Chun},
  booktitle={2025 IEEE Symposium on Security and Privacy (SP)},
  pages={4663--4681},
  year={2025},
  organization={IEEE}
}

@article{fei2025vocalcrypt,
  title={VocalCrypt: Novel Active Defense Against Deepfake Voice Based on Masking Effect},
  author={Fei, Qingyuan and Hou, Wenjie and Hai, Xuan and Liu, Xin},
  journal={arXiv preprint arXiv:2502.10329},
  year={2025}
}

@inproceedings{ASC1,
  title={Improving Acoustic Scene Classification in Low-Resource Conditions},
  author={Chen, Zhi and Shao, Yun-Fei and Ma, Yong and Wei, Mingsheng and Zhang, Le and Zhang, Wei-Qiang},
  booktitle={ICASSP 2025-2025 IEEE International Conference on Acoustics, Speech and Signal Processing (ICASSP)},
  pages={1--5},
  year={2025},
  organization={IEEE}
}

@article{li2025cloneshield,
  title={CloneShield: A Framework for Universal Perturbation Against Zero-Shot Voice Cloning},
  author={Li, Renyuan and Liang, Zhibo and Zhang, Haichuan and Shi, Tianyu and Cheng, Zhiyuan and Shi, Jia and Yang, Carl and Tang, Mingjie},
  journal={arXiv preprint arXiv:2505.19119},
  year={2025}
}

@inproceedings{feng2025enkidu,
  title={Enkidu: Universal Frequential Perturbation for Real-Time Audio Privacy Protection against Voice Deepfakes},
  author={Feng, Zhou and Chen, Jiahao and Zhou, Chunyi and Pu, Yuwen and Li, Qingming and Du, Tianyu and Ji, Shouling},
  booktitle={Proceedings of the 33rd ACM International Conference on Multimedia},
  pages={11638--11647},
  year={2025}
}

@article{kong2020panns,
  title={Panns: Large-scale pretrained audio neural networks for audio pattern recognition},
  author={Kong, Qiuqiang and Cao, Yin and Iqbal, Turab and Wang, Yuxuan and Wang, Wenwu and Plumbley, Mark D},
  journal={IEEE/ACM Transactions on Audio, Speech, and Language Processing},
  volume={28},
  pages={2880--2894},
  year={2020},
  publisher={IEEE}
}

@article{adam,
  title={Adam: A method for stochastic optimization},
  author={Kingma, Diederik P},
  journal={arXiv preprint arXiv:1412.6980},
  year={2014}
}

\end{document}